\documentclass[twocolumn,showpacs,amsmath,amssymb,aps]{revtex4}
\usepackage{graphicx}
\usepackage{epsfig}
\begin{document}

\title{  Mirror Energy Differences and nuclear structure in 1$f_{7/2}$ nuclei}
\author{
 F. Brandolini
}
\affiliation{
Dipartimento di Fisica dell'Universit\`a and INFN Sezione di Padova,
  I--35131 Padova, Italy
}


\begin{abstract}

Experimental mirror energy differences (MED) for nuclei lying in the middle and the second part of $1f_{7/2}$ shell are compared with shell model calculations in the $1f_{7/2}$ and in the full $pf$ configuration spaces, as well as with CSM calculations.
MED plots are fully consistent with the description which emerges from the interpretation of SM calculations: i.e. the band crossing of the gs band with high-$K$ bands in $A$=51 and 50 mirror pairs, as well as the seniority-3 quasi band-termination at $I^\pi$=19/2$^-$ in $A$=49. 
 Rotational alignment effects are excluded, contrarily to what even recently proposed.
 Some inconsistencies, as well as definition improprieties, are pointed out in the topical literature.
 
\end{abstract}



\maketitle

\section{Introduction}

 The interpretation of Coulomb displacement energies (CDE) or, more precisely, of the binding energy difference between isobaric analogue states (IAS) in isobaric multiplets, is a longstanding and still partly unresolved problem. Long ago, review articles discussed the contributing effects \cite{NS,Shl,Aue}.  The calculated CDE, considered as a Coulomb effect, were found to be systematically smaller than the experimental ones by 6-8 \%: Nolen-Schiffer (NS) anomaly \cite{NS}. The vibrational degree of freedom \cite{Shl2} and even nuclear isospin non conserving (INC) forces \cite{Shl,Aue} were considered as possible causes of the NS anomaly, but a general consent has been not achieved. 
Since then nuclear spectroscopy has been mainly interested in the measurement of Mirror Energy differences (MED), as they are believed to contain important nuclear-structure information \cite{BL}.  MED are the energy difference between analogue levels in level schemes of mirror nuclei: i.e. the CDE of mirror levels, referred to those of the respective ground state (gs). 

Most new data have been recently reviewed \cite{Ek1,BL}. The review article of Ref. \cite{BL}, much concentrated on 1$f_{7/2}$ nuclei, affirms, both in the abstract and conclusions, that one of its achievements was having shown MED to provide valuable and precise information on nuclear structure effects. However the article fails to mention the nuclear structure interpretations proposed in Ref. \cite{Br1} and further elaborated in Refs. \cite{Br2,BU}, which disagree with their own. Therefore all interpretations have to be discussed in more detail for a correct information. 
 
The present author has recently published, as first author, a review article on the nuclear structure of 1$f_{7/2}$ nuclei \cite{BU}, showing that most spectroscopic data in the middle and the second half of the shell are  well described  by   shell model (SM) calculations in the full $pf$ configuration space (CS), whose results  acquire a nuclear structure significance when compared with other models as rotor, particle-rotor and cranked Nilsson-Strutinsky (CNS). The description of 1$f_{7/2}$ nuclei turned out to be complex: there is evidence of nuclear deformation, which, however, coexists with properties of spherical nuclei, owing to the small number of active nucleons.
 It will be illustrated that those conclusions are proper tools to interpret the observed MED. Only levels of natural parity will be considered and several data that will be discussed are reviewed in  Ref. \cite{BL}.
 MED values  consist of various contributions \cite{NS,Shl,Aue,Ek1,BL}, but only those given by the SM Coulomb multipole matrix elements (i.e. among the valence protons) will be discussed in detail in sec.V, because it is the one which is more sensitive to nuclear structure properties. Other terms will be  briefly summarized to present a complete picture. 
 
The main task of the present paper is to extend the discussion of Ref. \cite{BU}, principally aimed at the nuclear structure information provided by dynamic (BM1), B(E2)) and static ($\mu, \, Q)$ electromagnetic (em) moments, to deepen the understanding of nuclear Coulomb energy and of its usage as a spectroscopic tool.
 
\section{MED in single-particle nuclei.}

Before examining specific 1$f_{7/2}$ nuclei within the frame of SM it is worth recalling relevant properties of the single particle (sp) case $A$=41 ($^{41}$Sc/$^{41}$Ca). The article by S. Shlomo \cite{Shl} will be principally followed in the present section. The contributions to CDE can be split into level-independent and level-dependent, in any case of the order or smaller than a hundred keV. Level-independent effects are various \cite{Shl}, but they do not affect MED.
Level-dependent sp contributions to be considered are: a) Coulomb Matrix Elements (CME), b) electromagnetic, c) short range correlation and other core-excitation, d) nuclear INC forces (for which no estimates are available).

 a) The contribution to MED of the sp CME is usually named radial, as it is obtained by integration over the radial wave-functions, and will be denoted later on with $V_{Cr}$. Sometimes it is evaluated assuming the same SM wave function in mirror nuclei for corresponding neutrons and protons. One should however consider also the second order correction related to the small Coulomb distortion of the potential well caused by the proton: it is called Thomas-Ehrman shift (TES) term and its size is at most of about a hundred keV for bound protons in low-{l} orbitals. In this paper $V_{Cr}$ includes TES, but in  literature the TES term is sometimes regarded as just the radial one \cite{Ek1}.
 
b) The electromagnetic corrections to the state energy are dominated by the spin-orbit contribution (EMSO) \cite{Ing} and its MED term will be denoted with $V_{Cls}$. There are also some smaller contributions as the orbit-orbit, the tensor and the proton density ones \cite{Shl}, which will be neglected here.
 The theoretical formula of EMSO, which applies to both protons and neutrons, is  
 
 \begin{equation}
 V_{ls}^{em}=(g_l-g_s){e \over 2m_N^2c^2}{dV_c \over rdr}l\cdot s
 \end{equation}

  Its quantitative validity  could not be checked because it is always concurrent with the radial term.

c) The short-range correlation is the part of core-excitation related to the closest distance between the valence nucleon and those of the core. It is larger for low orbital momentum and is usually neglected together with long-range correlation  \cite{Ek1,BL}, although they may be relevant \cite{Shl,Aue}.

The experimental MED value between the 2$p_{3/2}$ level and the 1$f_{7/2}$ gs is -227 keV. The sign is conventionally obtained by subtracting the level energy of the lower-Z mirror to the level energy of the upper-Z one.
The following contributions to MED were calculated in Ref. \cite{Shl} with a realistic Woods-Saxon (WS) potential: -260 keV (first order: -180 keV and TES: -80 keV), $V_{Cls}$: +70 keV, short range correlations: +30 keV. The order of magnitude is correct but the accuracy of each term is ambiguous. The Harmonic Oscillator (HO) potential must not be used because a  comparable value is obtained but with the wrong sign. This occurs because the asympotic radial behaviour is then gaussian, while in the case of a finite well is exponential. The same discrepancy occurs also for $A$=17 \cite{Shl}.

 It is known, however, that $^{40}$Ca is not a good core as it has about 30-40 \% mixing mainly consisting of deformed  4-hole configurations \cite{GG,Cau4}. The effect of such core-mixing on MED was not calculated. Empirical one- or two-body matrix elements extracted from $A$=41 and 42 are thus likely not reliable.

The sp mirror pair after the closure of the 1$f_{7/2}$ shell has $A$=57 ($^{57}$Cu/$^{57}$Ni) with 2$p_{3/2}$ gs configuration. This very important case has been studied both experimentally and theoretically in an article \cite{Tr}  which investigates in detail the relative importance of MED terms for excited levels using WS wave functions. Actually also $^{56}$Ni is not a good core, since SM show that it is mixed up at 50\%, but in this case the mixing is mainly with the upper orbitals of the $pf$ major shell CS. There the nucleus was described as a vibrator coupled with the sp states and the bare properties of the sp have been extracted from the analysis.
The  two lower excited level have configuration 1$f_{5/2}$ and 2$p_{1/2}$, respectively. The first excited level MED is of 260 keV, which is, as naively expected, about the opposite value of that of the 2$p_{3/2}$ level versus the 1$f_{7/2}$ gs in $A$=41. In fact $V_{Cr}$ depends principally on the orbital quantum number: $2p$ or $1f$. 
 The bare MED for the first excited level is 290 keV and the calculated bare contributions  are 190 keV for $V_{Cr}$ and 160 keV for $V_{Cls}$. This way their sum exceeds the bare MED by 60 keV. For the second excited 2$p_{1/2}$ level the experimental MED is -7 keV, its bare value is -50 keV, $V_{Cls}$ is +90 keV and $V_{Cr}$ is -120 keV. Thus, there is a qualitative agreement also in this case.

Before proceeding, it is necessary to remark that during the last ten years definitions have become rather confuse in literature.  Therefore, definitions will be refreshed and it will be commented in what they differ from  some others. 
$V_{Cr}$ is named $E_{ll}$  in Ref. \cite{BL}, where it was evaluated by means of an approximate formula, obtained in a HO basis \cite{Duf}. In this way $E_{ll}$= +150 keV was evaluated for the first excited level in $A$=41, which has the wrong sign, being a HO estimate. The sum of $E_{ll}$ and $V_{Cls}$  is about 230 keV which totally diverges from the experimental MED of -227 keV. The deviation of about -450 keV was interpreted at p. 532 of Ref. \cite{BL} as due to the difference in deformation  caused by the $2p_{3/2}$ and the $1f_{7/2}$ sp nucleons. Assuming a spherical gs, an average deformation parameter {$\beta \ge$  0.6 is estimated for the excited level using the classical formula:

 \begin{equation}
 \Delta E= {3 \over 25} {e^2 \over 4\pi\varepsilon_o} {Z^2 \over R} \epsilon^2  
 \end{equation}
 
where $\epsilon$ is the eccentricity parameter. This corresponds to superdeformation, which is clearly absurd as a large deformation is excluded in a sp state. The coefficient $E_{ll}$ has  even recently been used \cite{Dved} with misleading consequences.

\section{MED in multi-particle nuclei}

 The spherical shell model consists in approximating the interaction of each valence nucleons with the core with  a spherical potential, inclusive of the mean Coulomb field in the case of a valence proton. The rest of the binding energy is due the interaction among the valence nucleons. Assuming a two-body effective interaction the hamiltonian can be written:
 
 \begin{displaymath}
   H=\sum_i [K_i+V(r_i)+V_C(r_i)]+\sum_{ij}(V_{ij}+V_{Cij})
  \end{displaymath} 
Coulomb potential applies, obviously, only to protons. 
After separation of the terms depending on one or two indexes, one gets:
\begin{equation}
  H=H_m+H_M
 \end{equation}

\begin{figure*}[t]
\vspace*{-1.5 cm}
\epsfig{file=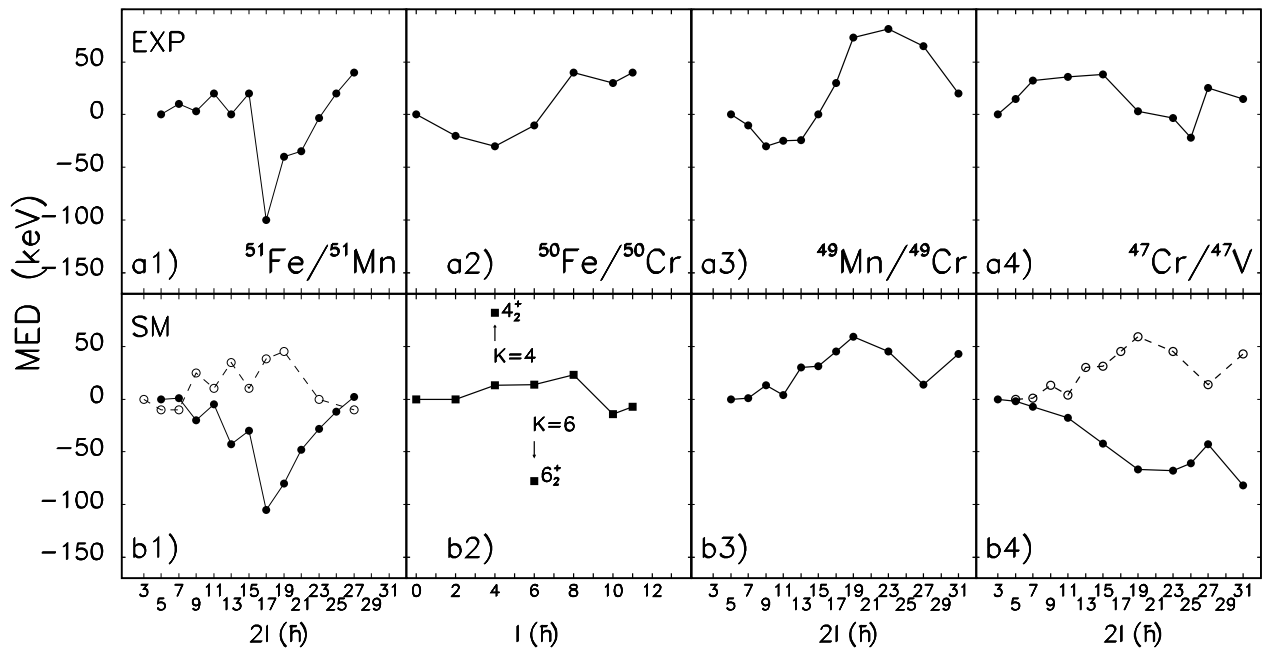,width=17.cm}
\caption{MED for $A$=51, 50, 49 and 47, a) Experimental [5], b) SM multipole matrix element contribution to CED  ($V_{CM}$).
 $V_{CM}$  for  $A$=45, conjugate of $A$=51, and for $A$=49, conjugate of  $A$=47, are plotted as open circles.}
\end{figure*}

 Thus, the state energy is the sum of the expectation values of the monopole and of the multipole hamiltonians.
 The monopole hamiltonian should  include all sp contributions for each orbital and type of nucleon: principally CME and EMSO. The experimental sp energies are usually adopted for diagonalization.

 
Monopole CME and EMSO contributions to the binding energy can be evaluated perturbatively after diagonalization, as they are much smaller than the total sp energies. As an eigenfunction is a linear combination of many basis functions, both terms are given by the  sp contributions in the different orbitals weighted on their occupancy. Owing to the definition of MED, $V_{Cr}$ and $V_{Cls}$ are the differences between the previous terms for protons and neutrons, respectively, referred to that of the respective gs.

The expectation values of the multipole Coulomb and of the nuclear contribution is expressed as a linear combination two-body CME, by means the angular momentum algebra.

A brief historical sketch is worth. In a pioneering work of Cameron et al. (1990) on $A$=49 \cite{Cam} MED variation along the yrast line was related to proton pair breaking and spin reorientation. It is known, in fact, that the mean mutual distance between two identical nucleons in the same orbital is minimum for I=0 and increases with spin. Thus Coulomb effect in a proton pair  varies with spin allowing to distinguish between proton and neutron pairs. Furthermore it was correctly observed that nuclei in the middle of the 1$f_{7/2}$ present evidence of rotational bands and hence of deformation, as it has been later beautifully shown \cite{Cau1}. Relying on these arguments the increasing of MED values along $I$ was qualitatively interpreted in terms of alignment of nucleon spin along the nuclear rotation axis (RAL) caused by Coriolis force, as elaborated in two subsequent theoretical papers based on cranked shell model (CSM) \cite{Sh1,Sh2} and an experimental one \cite{OL}.

The RAL interpretation was extended to MED plots of the yrast lines in $A$=47 \cite{Ben1}, 51 \cite{Ben2} and 50 \cite{Len}, which are considered rotational at low excitation energy. The experimental MED plots are reported in Figs. 1-a.
It turns out that MED values vary by less than 50 keV in the average.

 The alignment related to spin arrangement is crucial to explain the evolution of MED values along a yrast line. One should consider, however, that MED values provide only a global information which needs to be interpreted with the help of the known spectroscopy of the mirror pair. Other structure phenomena associated to alignment may occur: band-crossing with high-$K$ bands and `quasi' terminations in seniority subspaces. 

\section{Fit of experimental MED plots}

 Particular attention was given to the four yrast bands of  mirror pairs in Fig. 1-ax, because in the middle and the second half of the 1$f_{7/2}$ shell SM predictions are very good as illustrated in Ref. \cite{BU}. SM were performed there for natural parity states in the full $pf$ CS using the code ANTOINE \cite{Cau3}, freely available in Internet, with the nuclear interaction KB3G \cite{Pov1} and the Coulomb one. The obtained SM MED predictions, called $V_{CM}$, are reported in Fig. 1-bx and slightly differ in few cases from those reported in Ref. \cite{BL,Br1}. $V_{Cr}$ and $V_{Cls}$  are at most about  10\% of the single nucleon effects because the  terms for proton and neutron occupancy tend to elide.
  
  While SM reproduces very accurately standard spectroscopic features, delivers only qualitative MED plots. Empirical corrections beyond SM, most likely due to core effects, have been thus introduced.
 Two main deviations were first evidenced in Ref. \cite{Ben1}, when comparing $A$=49 and $A$=47 MED.
  A first deviation is that both $A$=49 and 47 MED plots lie higher than the SM predictions approaching band termination.
 Once this effect is corrected, a second deviation is that both plots are quenched at low spin toward the baseline with respect to SM predictions.
   The second effect was compensated by adjusting the low-spin two-body  CME ( Poves and Martinez-Pinedo in Ref. \cite{Ben1}).
  The first effect was accounted with a core deformation term, that will be called here $V_{Cd}$, whilst elsewhere $V_{Cr}^a$ \cite{Ek1}, $V_{Cr}$ \cite{BL} and $V_{Cm}$ \cite{Zuk}.  The prolate deformation was described using a configuration dependent cranked Nilsson potential \cite{Naz} yielding  $\beta\sim$0.22 at low spin  and nearly  zero at termination.  Therefore  $V_{Cd}$ increases with increasing spin in a similar way in $A$=47 and $A$=49.
 Essentially the same corrections were applied in Ref. \cite{Zuk}, where the fit was extended to $A$=51 and 50, but they were differently treated. 1) $V_{Cd}$ was guessed to be proportional to the $2p_{3/2}$ nucleon ( i.e. the sum of proton and neutron) occupancy. The term was there named $V_{Cm}$  and considered a monopole one, which is clearly incorrect in the sense of eq. 5. 2) It  was assumed that there is not a physical reason to modify  empirically the two-body CME, owing to the success of SM in this nuclear region. As a consequence  a further two-body force was introduced, with equivalent mathematical results as using the modified two-body CME. This new term was called $V_{BM}$ and was claimed to have a nuclear INC nature, without a direct proof. 
A reasonable fit was obtained in this way  for $A$=51, 50, 49, 47 as shown in detail in Refs. \cite{Ek1,BL}.  However a large discrepancy with experimental data occurs for the terminating level 31/2$^-$ in $A$=49 and for the 25/2$^-$ level in $A$=47, which will be discussed later.
 Anyhow, it must be stressed that the empirical corrections $V_{Cd}$ and $V_{BM}$ are not SM terms, as improperly  assumed in Ref. \cite{BL}.
 
 The same set of parameters reproduce approximately also the MED plots for $A$=42 and 54, which are nearly opposite and in both case the $I$=2 point is close to 0,  deviating from SM predictions by about 50 keV: `$I$=2 anomaly'. A cross-conjugation symmetry between 54 and 42 has been claimed \cite{Gad}, but such intuitive explanation conflicts with the  basically different structure of the two pairs. The impurity of the first 2$^+$ state in $^{42}$Ca is well known:
it consists of about 50 \%  of 2- and 4-hole configurations \cite{GG,Cau2}. Recent g-factors measurements \cite{Spe} confirm the large core mixing for the  2$^+$ levels in $^{42}$Ca and $^{44}$Ca, as the experimental values are positive, instead of largely negative. Core mixing becomes much smaller in $^{46}$Ca, thus approaching the doubly magic $^{48}$Ca. As previously noticed \cite{Ek2}, the $l$=3 spectroscopic factors for single-neutron pickup reactions leading to $^{42}$Ca are 0.86, 0.51, 0.88 for $I$=0, 2, 4, respectively \cite{End}, pointing out that core mixing is maximum in the 2$^+$ level.
  In this context it sounds rather odd that a nuclear INC force contribution, found neither in $sd$ nuclei nor for A$>$57 \cite{Ek1} may emerge for $A$=42 among rather uncontrolled effects. 
  
More reliable is the case $A$=54.  According to SM the 2$^+$ state of the two-hole gs band has more than 40 \% contribution of the upper part of the $pf$ CS. Still the terminating level of the gs band, which is commonly believed to be pure $f_{7/2}^{-n}$, is mixed by 40 \%. The same effect occurs in $^{52}$Fe and $^{53}$Fe \cite{BU}. The `2$^+$ anomaly' may be due to the asymmetric p-h core-excitation caused by valence protons and neutrons, respectively.

Owing to the uncertainty on $V_{Cr}$,  $V_{Cls}$, $V_{Cd}$ and $V_{BM}$ as well on core-mixing effects, the most reliable term in this nuclear region is $V_{CM}$.  With the help of basic models it will be shown to contain nuclear structure information.

\section{Nuclear structure and MED for some 1f$_{7/2}$ mirror pairs} 

\subsection{ $A$=51     ($^{51}$Fe/$^{51}$Mn)}

 Fig. 2 shows a partial level scheme of $^{51}$Mn, taken from Ref. \cite{BU}, with negative parity levels relevant for the present discussion. Similar schemes will be shown also for the other three low-Z members of the discussed mirror pairs. Levels with dominant 1$f_{7/2}$ configuration are displayed up to the terminating level $I^\pi$=27/2$^-$. As described in detail in Ref. \cite{BU} states evolve from prolate-collective to prolate-non collective in the Lund diagram. All  em moments are consistently predicted by SM. The band assignment is somewhat arbitrary above 21/2.

\begin{figure}
\vspace*{-0.5 cm}
 \includegraphics[width=6.8cm, clip]{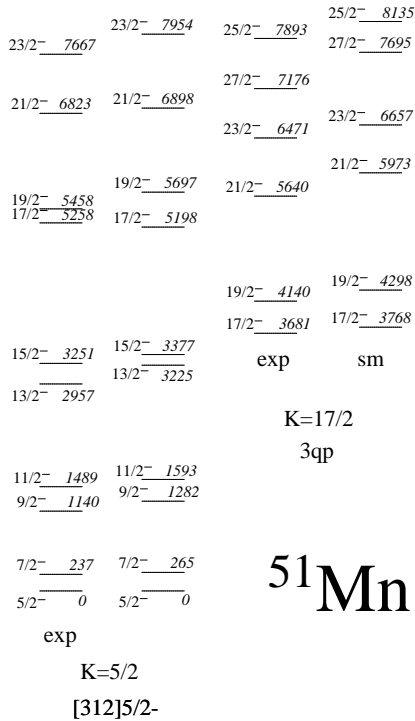}
 \protect\caption{Comparison of experimental negative parity levels in $^{51}$Mn with SM predictions [9].}
\end{figure}

\begin{figure}[t]
 \includegraphics[width=8.cm, clip]{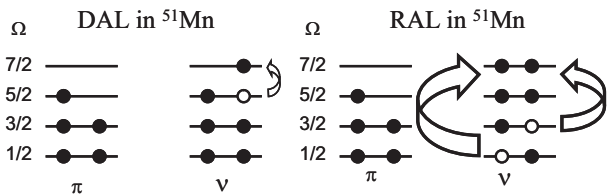} 
 \protect\caption{Comparison between DAL and RAL in $^{51}$Mn}
\end{figure}

Let us consider  the basic excitation features in $A$=51 in Fig. 3 in more detail.  In the prolate nucleus $^{51}$Mn the 6 neutrons fill the orbitals with $\Omega$=1/2, 3/2 and 5/2. The less expensive excitation is to break the up-most orbital $\Omega$=5/2 getting a neutron in $\Omega$=5/2 and one in $\Omega$=7/2. Coriolis coupling is weak for the active orbitals so that strong coupling i.e. deformation alignment (DAL) applies. One gets $K_N^\pi$=6$^+$ in the concordant case and  $K^\pi$=$17/2^-$ combining it with $K_P^\pi$=5/2$^-$, obtaining the Nilsson configuration $\nu$[303]7/2 $\nu$[312]5/2 $\pi$[312]5/2. Similar 3-qp bands occur in several $1f_{7/2}$ nuclei \cite{BU}
  
 The $K^\pi=17/2^-$ band is correctly reproduced by SM, which, in particular, predicts for the yrast $17/2^-$ level an electric quadrupole moment $Q$=+56.3 efm$^2$, corresponding to
 a deformation parameter of $\beta \approx$0.22  for $K^\pi$ =17/2$^-$, somewhat smaller, as reasonable, than that of low-lying levels ($\beta \approx$0.25). The value of $Q$ is otherwise totally inconsistent with the negative value predicted for $K^\pi\approx 5/2^-$ as it would be in case of a sudden RAL.

The MED plot in $A$=51 mirror pair (Fig. 1-a1) is peculiar as it experiences a sudden drop at $I^\pi=17/2^-$  \cite{Ek3,Ben2}. The theoretical $V_{CM}$ plot in Fig. 1-b1 reproduces the behavior rather well, so that the sudden change of MED at $I$=17/2 is immediately  related with the band-crossing with a high-$K$ band. 
 Band-crossings of the gs band with a high-$K$ 3-qp bands are only predicted and observed in the second half of the 1$f_{7/2}$ shell, owing to the higher value of $\Omega$ involved, as for example in $^{51}$Cr \cite{BU}. 

The previous discussion shows that the DAL description agrees with the known spectroscopy, while the RAL one does not. It has to be remarked, however, that a RAL interpretation of the sudden drop has been given even recently. Ref. \cite{War} comments at p. 313: ``such CED effect is associated with a sudden rotational alignment: Coriolis force breaks the coupling of a pair of identical nucleons whose angular momentum vectors align suddenly from $I$=0 to 6, the maximum allowed for a pair in this shell''. The same description is given in Ref. \cite{Ben2} and  \cite{BL}, where the experimental plot is interpreted according to the CSM calculations of Ref. \cite{Sh2}, just a continuation of \cite{Sh1}.

 \begin{figure}[t]
\begin{center}
\hspace*{-0.5 cm}
\includegraphics[width=7.5cm, clip]{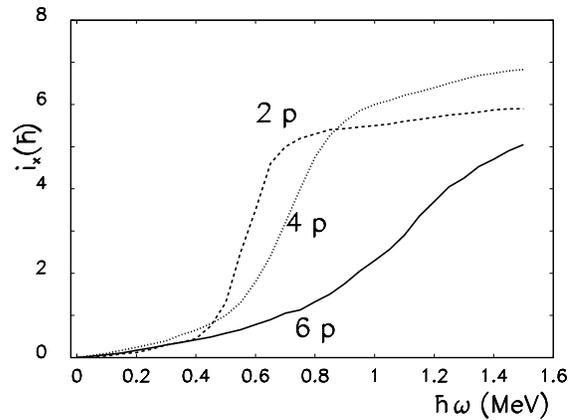}
\end{center}
\vspace{0.5 cm}
\caption{Rotationally aligned proton-spin $i_x$ versus the rotational frequency, for 2, 4, 6 protons in the 1$f_{7/2}$ shell, assuming $\beta$=0.20. (taken from Ref. 20)}
\end{figure}

\begin{figure*}[t]
\vspace*{-1.5 cm}
\epsfig{file=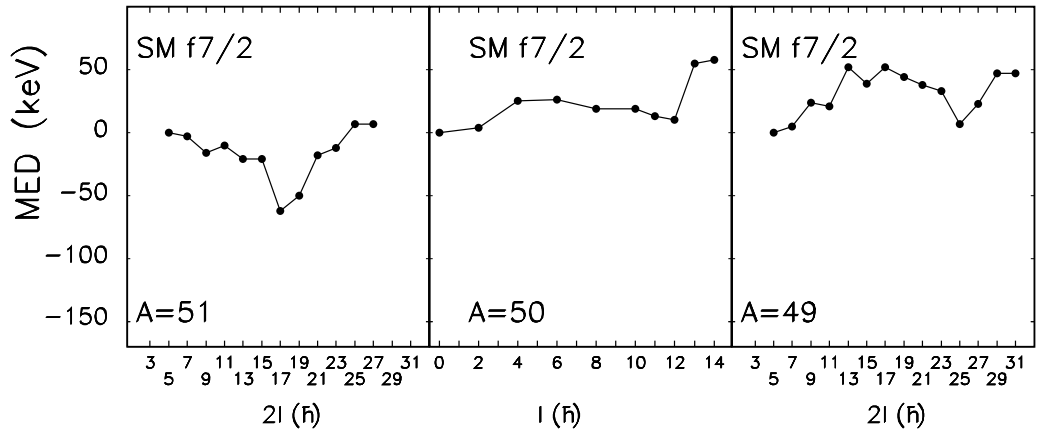,width=14.cm}
\caption{MED predictions along the yrast line in the pure $1f_{7/2}$ configuration space for $A$=51, 50 and 49. Exactly the opposite predictions apply to $A$=45, 46 and 47, respectively, because of p-h conjugation symmetry.}
\end{figure*}

In order to definitively reject the later interpretation one will show that even the invoked CSM calculations  \cite{Sh2} exclude RAL occurrence in $A$=51.
 For this purpose the alignment plots along a perpendicular to the symmetry axis are shown  in Fig. 4 for 2, 4 and 6 protons. The deformation assumed there is $\beta$=0.2. 
 Let us briefly summarize the most important features illustrated in Ref. \cite{Sh2}. In the case of odd A mirror pair, as the present one, if a nucleus has an even number of protons, the mirror one has an odd number. In the same reference it is observed that the main multipole features are provided by the even proton mirror, because the alignment of an odd number of protons is rather flat both for 3 protons ($A$=47) and 5 protons ($A$=49), being partially blocked by the unpaired proton.
 In both $A$=47 and 49 the four-proton system prevails, leading to a positive MED in $A$=49 and to a negative one in $A$=47.

   The predicted alignment becomes very slow  in the 6-proton case which applies to $A$=51. The  assumed deformation is somewhat smaller than the experimental value 0.24 estimated for low levels in $A$=51 \cite{BU}, so that the alignment would be even slower in the real case, because a larger deformation hinders RAL.
Looking at the experimental level scheme in Fig. 2, the alignment of the yrast 17/2$^-$  occurs suddenly at about $h\omega \approx$ 1 MeV, where the proton alignment is just slowly going on, as shown in Fig. 3. The conclusion is drawn that the model fails in reproducing the experimental MED plot for $A$=51. It is surprising that a misinterpretation \cite{Ben2} has led to affirm  the contrary for such a long time.

 As shown in Fig. 3, neutron pair RAL with $i_x$=6 is not realistic in $^{51}$Mn because one would need to make a hole in the low-lying $\Omega$=1/2 and 3/2 orbitals simultaneously filling the $\Omega$=7/2 orbital.  In a  simplified picture the order of magnitude both of orbital separation and of pairing energy is about 2 MeV, so that  an excitation energy several times bigger is required. It becomes even energetically favored to make RAL in proton space. This  is reflected in the very slow alignment of Fig. 4 and in very small predicted MED \cite{Sh2}. In Fig. 3 the anti-pairing action of Coriolis force is not represented but could  lead to an energy saving of about 2 MeV.
 
 RAL bands are frequently observed in heavier nuclei and can be also superdeformed, but in that case they are intruder and not generated  below the Fermi level as in the present case. Moreover in the present case rotational motion is very frail because it is built on a limited number of valence nucleons, so that an early band termination occurs. 
 The band in $^{51}$Mn is prolate-rotational only at low excitation, it becomes largely triaxial and finally prolate non collective approaching termination   \cite{BU,Ju3}.
 Triaxiality may explain the signature staggering in $A$=51 MED.
 The assumption of a fixed $\beta$ along the band \cite{Sh2} is thus not realistic.
 
 Fig. 5 shows  MED predictions for $A$=51, 50 and 49  in the pure $1f_{7/2}$ CS. For $A$=51 the figure resembles $V_{CM}$, even if  with a much smaller dip, which seems to be a seniority 3 effect. If seniority is a quasi good quantum number, the states 19/2 and 17/2 in $^{51}$Mn can  only be obtained coupling $I_N$=6 and $I_P$=7/2.
The mixing with the $2p_{3/2}$ orbital makes the two states rotational, increasing $Q$($I$=17/2) and the MED dip by more than a factor of two.

In fig. 1-b1 the SM MED plot for $A$=45 is also reported.
$A$=45 is particle-hole (p-h) conjugate of $A$=51 so that the opposite MED is predicted in a pure $1f_{7/2}$ CS, because of the invariance under p-h conjugation \cite{BMZ}. In fact the $A$=51 pair is described by 6p-5n/5p-6n, which becomes 2p-3n/3p-2n  by p-h conjugation, which has opposite MED than $A$=45. 
 According to Fig. 4 the alignment is expected to be very steep in the 2-proton case, since the two protons belonging to the [330]1/2 orbital are easily split as Coriolis force is rather effective in this case. If CSM is correct, also SM should predict a sudden step, while it is not so.  The 2-proton plot is meaningless, however, because if the two protons are completely decoupled they do not cooperate to the rotation but $^{43}$Ca does not rotate, so that the assumed deformation parameter $\beta$=0.2 is unphysical.

One has to conclude that CSM cannot describe MED for both $A$=51 and $A$=45. RAL cannot occur also in A=46 for the same reason as for A=45.

\begin{figure}[t]
 \includegraphics[width=8.cm, clip]{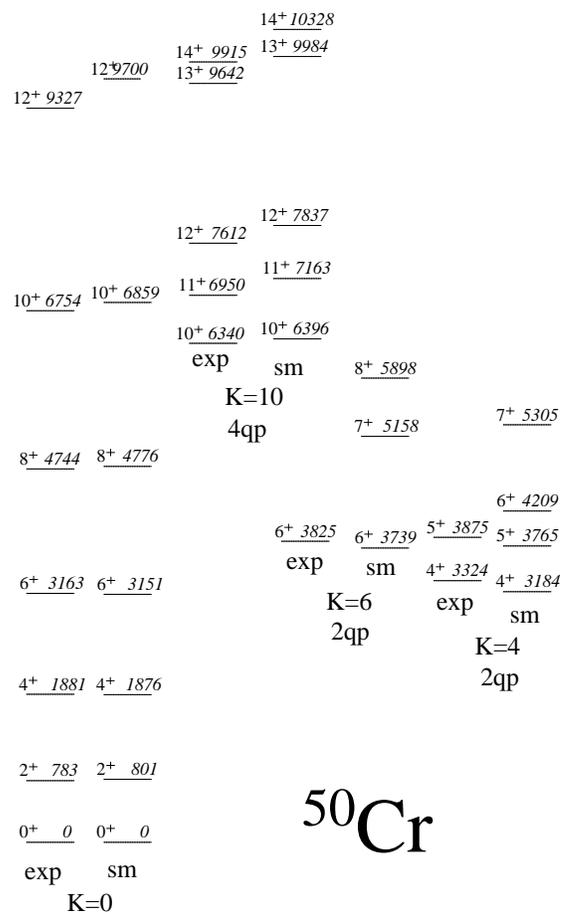} 
 \protect\caption{Comparison of experimental negative parity levels in $^{50}$Cr with SM predictions [9].}
\end{figure}

\subsection{ $A$=50 ($^{50}$Fe/$^{50}$Cr)}

\begin{figure}[t]
 \includegraphics[width=8.cm, clip]{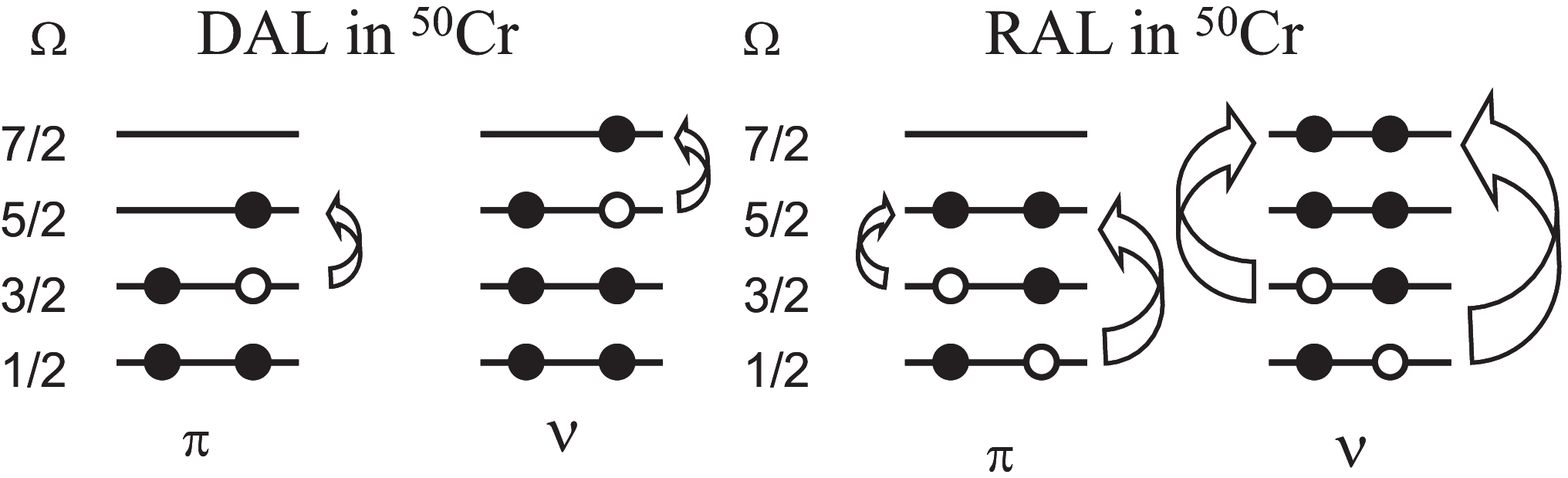} 
 \protect\caption{Comparison between DAL and RAL in $^{50}$Cr}
\end{figure}

 The $^{50}$Cr experimental level scheme, extracted from Ref. \cite{BU}, is compared with SM predictions in Fig. 6.  This case was discussed in Ref. \cite{Br1}.
The sideband with $K$= 4 is obtained with the excitation of a proton and has configuration   $\pi$[321]3/2 $\pi$[312]5/2. The sideband with $K$=6 is obtained with the excitation of a neutron and has configuration   $\nu$[303]7/2 $\nu$[312]5/2.      
The main features in $^{50}$Cr, predicted by SM and in agreement with the experimental level scheme, are:  with increasing excitation energy first a $K$=4 band is formed breaking a proton pair, in alternative, little afterward a $K$=6 band is made breaking a neutron pair. Thus  Nilsson orbitals are nearly equidistant. The two band heads have SM g-factor 1.2 and -0.2, pointing on their proton and neutron nature, respectively. Therefore these side-bands are  predicted to have an opposite MED behavior as it is shown in Fig. 1-b2. 

Breaking both pairs simultaneously one gets the yrast 4-qp band with $K$=10, which gives rise to the MED dip at $I$=10 along the yrast line in Fig. 1-a2.
 In fact the MED effect of  $K=$6  prevails, as shown in Fig. 1-b2.  SM predicts a larger dip, but calculations are less precise in case of band-crossing.

     SM electric quadrupole moment $Q$ becomes suddenly positive for $I$=10 along the yrast line. 
       RAL description of MED, which implies a negative $Q$ for $I$=10,  was shown to be inconsistent with SM calculations, with similar arguments as in $^{51}$Mn \cite{BU}. Moreover the yrast 11$^+$ level is consistent only with a DAL description.
  
  Fig. 7 shows that RAL would occur in the proton space, although  much hindered with respect to DAL. RAL of neutrons would occur at much higher energies and certainly where the deformation is exhausted, in  contrast with the statement (p. 544 of Ref. \cite{BL}) that they should occur at similar excitation energies.

One sees in Fig. 5 that positive MED are predicted at low spin also in pure $1f_{7/2}$ CS. The main difference is that the $I$=10 value of $V_{CM}$ is drastically depressed.

\begin{figure}[h]
  \includegraphics[width=6.4cm, height=12cm, clip]{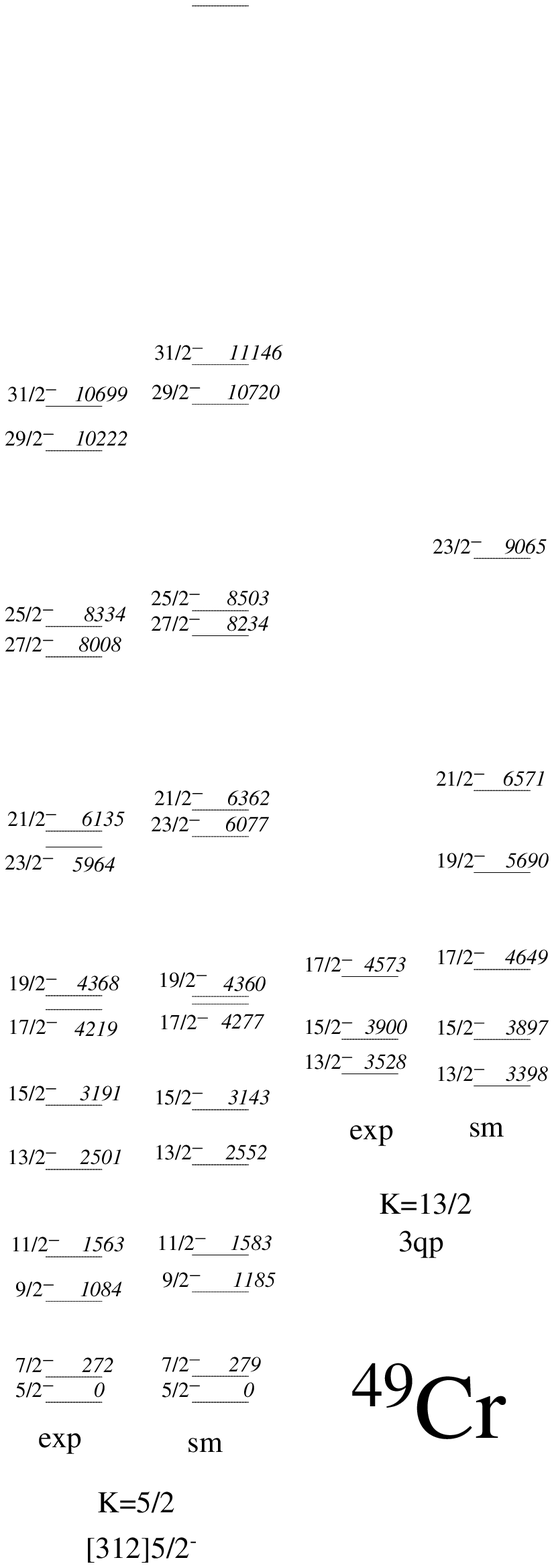}
 \protect\caption{Comparison of experimental negative parity levels in $^{49}$Cr with SM predictions [9].}
\end{figure}

\subsection{ $A$=49  ($^{49}$Mn/$^{49}$Cr)}

 MED in the $A$=49 nuclei was the first case studied in detail \cite{Cam,OL}. 
 Experimental and SM points are reported in Figs. 1-a3 and 1-b3, respectively.

  In Ref. \cite{Br3} a nearly full spectroscopy in $^{49}$Cr has been made  up to more than 4 MeV of excitation. All SM levels have been calculated in the same energy range and nearly each of them has an experimental counterpart close in energy.
  In Fig. 8  only the relevant negative parity levels in $^{49}$Cr are compared to SM predictions. The calculated em properties agree with those expected for the gs band. The deduced deformation parameter for low spins is $\beta$=0.27.
  In analogy with the 3-qp band in $^{51}$Mn, a sideband $K^{\pi}$=13/2$^-$ occurs,  which is obtained by lifting one proton from the orbital [321]3/2 to the [312]5/2, as shown in Fig. 9, and coupling the three unpaired particles to the maximum $K$ value. The value of $Q$  of the $K^\pi$=13/2$^-$ band head is 51.2 efm$^2$, which corresponds to $\beta$=0.24, just a bit smaller than that of gs band, as reasonable. The back-bending at $I^\pi$=19/2$^-$  is well reproduced by SM. Back-bending is related  only to this level and thus only to the $\alpha$=-1/2 signature.

  Ref. 5 writes  about the back-bending at p. 535: ``At around $I^\pi$=19/2$^-$ a rotational alignment of a pair of protons occurs in $^{49}$Cr, the alignment of neutrons being blocked by the unpaired 1$f_{7/2}$ neutron. As the protons align from I = 0 to the maximum allowed I = 6''.
The back-bending at $I$=19/2 is, however, not due to RAL, which is energetically unfavored with respect to DAL, as shown in Fig. 9.

\begin{figure}[h]
 \includegraphics[width=8.cm, clip]{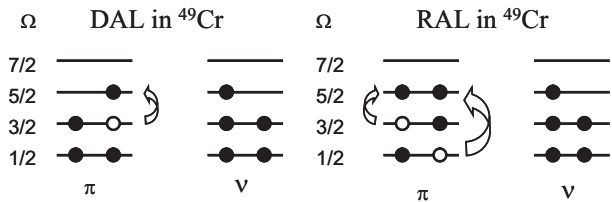} 
 \protect\caption{Comparison between DAL and RAL in $^{49}$Cr}
\vspace*{-.5cm}
\end{figure}

The $A$=49 MED plot of Fig. 5, predicted in the pure $1f_{7/2}$ CS, is largely positive and resembles $V_{CM}$.  Protons and neutrons align considerably and differently also in spherical nuclei, eradicating an old prejudice, so that there was not a physical motivation for the first suggestion of rotational alignment \cite{Sh1}. Moreover the nearly opposite behavior of $V_{CM}$ for $A$=49 and 47 is a consequence of p-h symmetry and not of RAL. It is astonishing that such a simple observation was not made so far. Beside this fundamental remark there are also experimental arguments against RAL:

a) The lowest level of a RAL band in an odd-N 1$f_{7/2}$  nucleus can be 19/2$^-$ only if $K_N=1/2$, because the neutron spin is then also oriented along the rotational axis. In the present case $K_N=5/2$ so that the back-bending is predicted at $I$=15/2 or 17/2.

b) A condition of proton pair RAL  to occur along the yrast line is the existence of the RAL band $\nu$[330]1/2 which should give rise to a decoupled band 7/2, 11/2, ...,which is neither predicted by SM nor observed below 4 MeV. 


c) If the yrast 19/2$^-$ level is rotational aligned, the corresponding level belonging to the gs band should be few hundred keV above. The yrare 19/2$^-$ level is, however, predicted by SM  1.3 MeV above the yrast one, but it belongs to the 3-qp band  $K=13/2^-$.

d) In Ref \cite{Br4} the B(E2) values show a regular pattern in accord with SM predictions. Since the back-bending appears only in one level of  signature $\alpha$=-1/2 , the crossing with a RAL band would occur with little mixing, in which case a drop of B(E2) would be observed.

Owing to the difficulties encountered by a RAL description, nearly ten years ago the present author described the back-bending at $I$=19/2  as due to a quasi band-termination in the seniority-3 subspace \cite{BrS}. This is consistent with the fact that exactly in correspondence a maximum appears in $V_{CM}$ (Fig. 1-b3). In conjunction, the back-bending at $I$=12 in $^{48}$Cr was described as due to a seniority-4 `quasi' band termination. The same description of the two back-bending was proposed in Refs. \cite{Ju1,Ju2} on the basis of a detailed SM and CNS analysis and discussed in Ref. \cite{BU}. These interpretations are reasonable considering that the two nuclei may be obtained by removing a $\alpha$-particle from $^{53}$Fe and $^{52}$Fe, which both exhibit energy inversion, with isomerism, at the gs band-termination with seniority 3 and 4, respectively. 

\subsection{ $A$=47  ($^{47}$Cr/$^{47}$V)}

The level scheme of $^{47}$V is shown in Fig. 10, for the relevant levels \cite{BU}. Experimental MED values are reported in Fig. 1-a4. The yrast 17/2$^-$ level is not known, as well as most levels of the 3-qp band.

The $A$=47 pair is p-h conjugate of the $A$=49 one and thus the gs-band also terminates at 31/2$^-$. The deformation parameter $\beta$ is 0.26 at low spin, a bit smaller than for $A$=49. This case is usually discussed together with the $A$=49 one. The calculated SM $V_{CM}$ term is, in fact, nearly opposite to that for $A$=49 as shown in Fig. 1-b4,  but it is somewhat shifted toward higher spins.  p-h symmetry applies fairly well to states close to termination because the configuration mixing is only little more than 10 \% \cite{BU}. Such levels in $A$=47 follow the SM prediction, while in $A$=49 the 31/2 strongly deviates. Since cross-conjugation should be approximately valid the terminating 31/2 point in $^{49}$Cr is most likely wrong. This is confirmed by explicit calculations in  pure $1f_{7/2}$ CS reported in Fig. 11.

\begin{figure}[b]
  \includegraphics[width=7.2cm, clip]{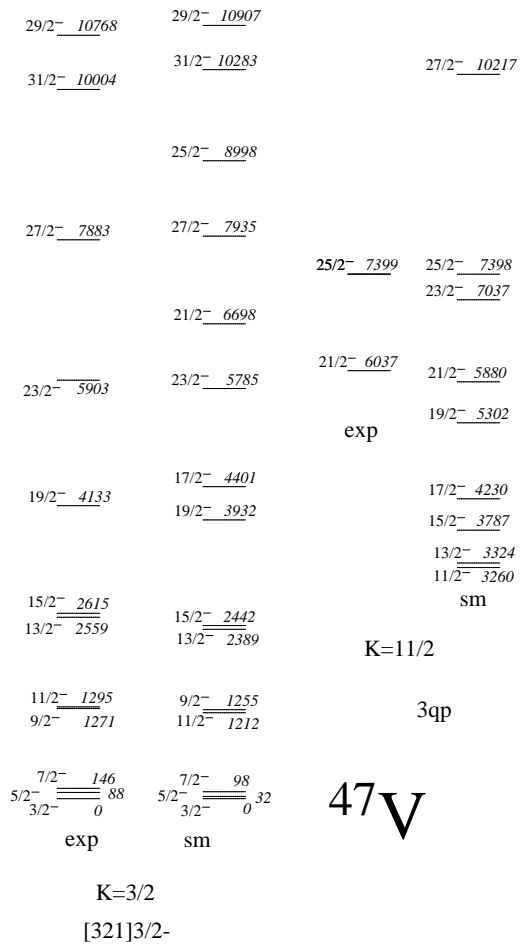}
 \protect\caption{Comparison of experimental negative parity levels in $^{47}$V with SM predictions [9].}
\end{figure}

 It is remarkable that  $V_{CM}$ has the largest deviation from 0 at termination, contradicting the comment at p. 536 of Ref. \cite{BL}: "MED reduces at high spins". Large deviations are predicted also for $A$=50 and 49.  


{
\begin{figure}[t]
 \includegraphics[width=7.5cm, clip]{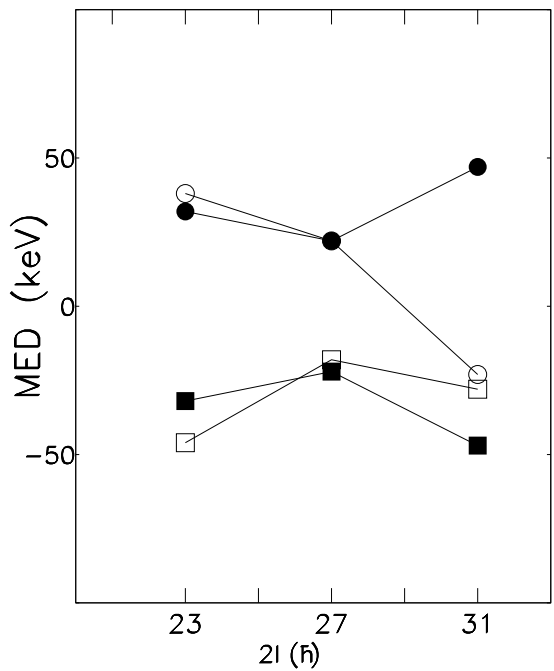} 
 \protect\caption{Open cirles: $A$=49 experimental MED of the $\alpha$=-1/2 signature approaching band termination ($V_{Cd}$=45 keV is subtracted). Open squares: the same for $A$=47.  Full circle $A$=49  MED calculated in the pure $1f_{7/2}$ configuration. Full square: the same for $A$=47.}
\end{figure}  

 In $^{47}$V  there is no evidence of back-bending at $I^\pi$=19/2$^-$ along with the $\alpha$=-1/2 signature sequence, as it was in the case of $A$=49. This fact has been recently examined in Ref. \cite{Ju3}, which explains the different behavior  with the evidence that, when calculating the pairing energy along that signature, a maximum is predicted at 19/2 only for $A$= 53, 49 and 45, i.e. nuclei differing by an $\alpha$-particle. This is likely related to  the absence of a clear minimum at 19/2 in the $A$=47 $V_{CM}$  plot.

  Finally, the experimental discontinuity of the MED plot at $I$=25/2 in $^{47}$V is likely related to the band crossing predicted by SM in Fig. 6 and thus to a different structure.

\section{Conclusions}

The present work deals mainly with the interpretation  of MED in mirror pairs with $A$=51, 50, 49 and 47.
While most properties of low-lying natural parity levels of these 1$f_{7/2}$ nuclei are very well described by SM calculations in the full pf CS \cite{BU}, SM calculated MED agree with experimental ones only qualitatively, so that two corrective terms, beyond SM, were introduced ten years ago \cite{Ben1}.

The sp MED terms are the radial $V_{Cr}$ and the electromagnetic $V_{Cls}$.
 The coefficient $E_{ll}$ in Refs. \cite{Duf,BL}, corresponding to  $V_{Cr}$ in present article, has the wrong sign, being based on HO  wavefunctions, rather than on the realistic WS one \cite{Shl}. This has been shown here to lead to absurd consequences  and to impair the MED interpretation of Ref. \cite{BL}. 
 The em spin-orbit $V_{Cls}$ term cannot be measured separately from the radial one.
Both terms  are anyhow of secondary importance in the examined cases \cite{Ek1}, because the differences  between proton and neutron  contributions are small.
 
The two empirical corrective terms are named $V_{Cd}$ and $V_{BM}$. 
 $V_{Cd}$ is related to the deformation of the core \cite{Ben1}. It gives rise to positive contribution for all MED plots approaching termination. It is elsewhere considered a monopole term \cite{Zuk,BL}, but this is incorrect in the SM terminology as it depends there on the sum of neutron and proton 2$p_{3/2}$ occupancy.
The $V_{BM}$ term accounts of the `$I$=2 anomaly' in mentioned mirror pairs and more specifically for A=54. Its microscopic origin is not yet understood: it may be a core-polarization effect, but also a nuclear INC force has been invoked \cite{Zuk,BL}.

The multipole CME give rise to the MED $V_{CM}$ term. Few inconsistencies in literature have been here pointed out also in this case. The statement, that  MED should reduce to small values approaching termination \cite{OL,BL}, is not correct: on the contrary they get the maximum deviation in $A$=47, 49 and 50.
 The experimental 31/2$^-$ point in $A$=49 \cite{OL,BL} is likely wrong, because it strongly violates the p-h conjugation symmetry.

The structural interpretation has been based here on the $V_{CM}$ term as it reproduces most  features of experimental MED plots. 
  According to the SM analysis  the sudden drop of MED at $I$=17/2 in $A$=51 is attributed to the crossing of the gs band with a 3-qp high-$K$ band. In a similar way the MED dip  at $I$=10 in $A$=50 is due to  the band-crossing with a 4-qp high-$K$ band. 
For $A$=49  SM confirms  a `quasi' band-termination in a seniority-3 subspace at $I$=19/2, which is associated to back-bending and to a secondary maximum  in the $V_{CM}$ plot.

The present nuclear structure interpretations of MED plots in $A$=51, 50, 49 and 47 disagree with the one reported in the review article of Ref. \cite{BL}, where RAL along the yrast line was proposed for all cases.  SM in the pure 1$f_{7/2}$ CS predicts in $A$=49 a MED plot rather similar to $V_{CM}$, showing that the early proposal of  RAL \cite{Sh1} was not physically motivated.
 Several experimental and theoretical arguments have been put forward, showing that RAL does not occur in the examined cases.


\begin{thebibliography}{200}

\bibitem{NS} J.A. Nolen and J.P. Schiffer, Annu. Rev. Nucl. Sci. $\bf 19$,  471 (1969). 

\bibitem{Shl} S. Shlomo, Rep. Prog. Phys., vol. $\bf 41$,  957 (1978).

\bibitem{Aue} N. Auerbach, Phys. Rep. $\bf 98$,  273 (1983).

\bibitem{Shl2} S. Shlomo and W.G. Love, Phys. Scripta 26, 280 (1982).

\bibitem{BL} M.A. Bentley and S.M. Lenzi, Progr. Part. Nucl. Phys. $\bf 59$, 497 (2007).

\bibitem{Ek1} J. Ekman, C. Fahlander and D. Rudolph, Mod. Phys. Lett. A $\bf 20$, 2977 (2005).

\bibitem{Br1} F. Brandolini {\it et al.}, Phys. Rev. C $\bf 66$, 021302 (2002).

\bibitem{Br2} F. Brandolini, Eur. Phys. J. A $\bf 20$,  139 (2004).

\bibitem{BU} F. Brandolini and C.A. Ur, Phys. Rev. C $\bf 71$,  054316 (2005), nucl-th/0407096 (2004).

\bibitem{Ing} D.R. Inglis, Phys. Rev. $\bf 82$,  181 (1951).

\bibitem{GG} W.J. Gerace and A.M. Green, Nucl. Phys. A $\bf 93$,  110 (1967).

\bibitem{Cau4} E. Caurier et al., Phys. Rev. C 75, 054317 (2007).

\bibitem{Tr} L. Trache {\it et al.}, Phys. Rev. C $\bf 54$,  2361 (1996).

\bibitem{Duf} J. Duflo and A.P. Zuker, Phys. Rev. C $\bf 66$, 051304 (2002).

\bibitem{Dved} F. Della Vedova et al., Phys. Rev. C $\bf 75$, 034317 (2007).

\bibitem{Zuk} A. P. Zuker {\it et al.}, Phys. Rev. Lett. $\bf 89$, 142502 (2002).

\bibitem{Cam} J.A. Cameron {\it et al.}, Phys. Lett. B $\bf 235$ (1990). 

\bibitem{Cau1} E. Caurier {\it et al.}, Phys. Rev. C $\bf 50$, 225 (1994).

\bibitem{Sh1} J.A. Sheikh {\it et al.}, Phys. Lett. B $\bf 252$, 314 (1990).

\bibitem{Sh2} J.A. Sheikh, D.D. Warner and P. van Isacker, Phys. Lett. B $\bf 443$, 16 (1998).

\bibitem{OL} C.D. O'Leary {\it et al.}, Phys. Rev. Lett. $\bf 79$ , 4349 (1997) .

\bibitem{Ben1} M.A. Bentley {\it et al.}, Phys. Lett. B $\bf 437$,  243 (1998), ERRATUM, Phys. Lett. B $\bf 451$,  445 (1999).

\bibitem{Ben2} M.A. Bentley {\it et al.}, Phys. Rev. C $\bf 62$,  051303 (2000).

\bibitem{Len} S.M. Lenzi {\it et al.},  Phys. Rev. Lett. $\bf 87$,  122501 (2001).


\bibitem{Cau3} E. Caurier and F. Nowacki, Acta Phys. Pol. {\bf 30}, 705 (1999).

\bibitem{Pov1} A. Poves {\it et al.}, Nucl. Phys. A $694$, 157 (2001).

\bibitem{Naz} A. Afanasjev and I. Ragnarsson, Nucl. Phys. A 591, 387 (1995).

\bibitem{Gad} A. Gadea {\it et al.}, Phys. Rev. Lett $\bf 97$, 152501 (2006).


\bibitem{Cau2} E. Caurier {\it et al.}, Phys. Lett. B $\bf 522$, 240 (2001).

\bibitem{Spe} K.-H. Speidel {\it et al.},  Phys. Rev. C $\bf 68$, 061302  (2003) and refs therein.

\bibitem{Ek2} J. Ekman, Ph.D thesis. Lund University, Sweden, 2004.

\bibitem{End} P.M. Endt and C. van der Leun, Nucl. Phys. A $\bf 310$,  1 (1978).


\bibitem{Ek3} J. Ekman {\it et al.}, Eur. Phys. J. A $\bf 9$,  13 (2000).

\bibitem{War} D.D. Warner, M.A. Bentley and P. van Isacker, Nat. Phys. $\bf 2$, 311  (2006). 


\bibitem{BMZ} J.D. Mc Cullen, B.F. Bayman and L. Zamick, Phys. Rev. 134, B515 (1964).

\bibitem{Ju3} A. Juodagalvis, I. Ragnarsson and S. Aberg, Phys. Rev. C $\bf 73$,  044327 (2006). 

\bibitem{Zam} L. Zamick and D.C. Zheng, Phys. Rev. C $\bf 54$, 956 (1996).

\bibitem{Br3} F. Brandolini {\it et al.}, Phys. Rev. C $\bf 73$, 024313 (2006).

\bibitem{Br4} F. Brandolini {\it et al.}, Nucl. Phys. A $\bf 693$, 517 (2001).

\bibitem{BrS} F. Brandolini in `Experimental Nuclear Physics in
Europe Facing the Next Millenium' Seville, 21-26 June 1999. Eds. B. Rubio, M.
Lozano and W. Gelletly. AIP Conference Proceedings 495, p. 189.

\bibitem{Ju1} A. Juodagalvis and S. \AA berg, Phys. Lett. B $ 428$, 227 (1998).

\bibitem{Ju2}  A.~Juodagalvis, I.~Ragnarsson and S.~Aberg, Phys. Lett.B {\bf 477}, 66 (2000).
 





\end{thebibliography}
\end{document}